\documentclass[aps,prl,twocolumn,showpacs,superscriptaddress,groupedaddress,graphics]{revtex4}
\usepackage{graphicx}
\usepackage{dcolumn}
\usepackage{bm}
\usepackage{amssymb}
\usepackage{mwe}
\usepackage{comment}
\usepackage{xcolor}

\begin{document}

\title{First prototype of a biaxially textured YBa$_{2}$Cu$_{3}$O$_{7-x}$ microwave cavity\\in a high magnetic field for dark matter axion search}
\affiliation{Center for Axion and Precision Physics Research, Institute for Basic Science, \\Daejeon 34051, Republic of Korea}
\author{Danho Ahn} \affiliation{Department of Physics, Korea Advanced Institute of Science and Technology (KAIST), \\Daejeon 34141, Republic of Korea} \affiliation{Center for Axion and Precision Physics Research, Institute for Basic Science, \\Daejeon 34051, Republic of Korea}
\author{Ohjoon Kwon} \affiliation{Center for Axion and Precision Physics Research, Institute for Basic Science, \\Daejeon 34051, Republic of Korea}
\author{Woohyun Chung} \email[Corresponding author.\\]{gnuhcw@ibs.re.kr} \affiliation{Center for Axion and Precision Physics Research, Institute for Basic Science, \\Daejeon 34051, Republic of Korea}
\author{Wonjun Jang} \altaffiliation{Present address: Samsung Advanced Institute of Technology, Suwon, 16678, Republic of Korea.} \affiliation{Center for Quantum Nanoscience, Institute for Basic Science, \\Seoul 33760, Republic of Korea}
\author{Doyu Lee} \altaffiliation{Present address: Samsung Electronics, Hwaseong, 18448, Republic of Korea.} \affiliation{Center for Axion and Precision Physics Research, Institute for Basic Science, \\Daejeon 34051, Republic of Korea}
\author{Jhinhwan Lee} \affiliation{Center for Artificial Low Dimensional Electronic Systems, Institute for Basic Science, \\Pohang 37673, Republic of Korea}
\author{Sung Woo Youn} \affiliation{Center for Axion and Precision Physics Research, Institute for Basic Science, \\Daejeon 34051, Republic of Korea}
\author{HeeSu Byun} \affiliation{Center for Axion and Precision Physics Research, Institute for Basic Science, \\Daejeon 34051, Republic of Korea}
\author{Dojun Youm} \affiliation{Department of Physics, Korea Advanced Institute of Science and Technology (KAIST), \\Daejeon 34141, Republic of Korea}
\author{Yannis K. Semertzidis} \affiliation{Center for Axion and Precision Physics Research, Institute for Basic Science, \\Daejeon 34051, Republic of Korea} \affiliation{Department of Physics, Korea Advanced Institute of Science and Technology (KAIST), \\Daejeon 34141, Republic of Korea}
\date{\today}

\begin{abstract}
A high-quality factor microwave resonator in the presence of a strong magnetic field could have a wide range of applications, such as axion dark matter searches where the two aspects must coexist to enhance the experimental sensitivity.
We introduce a polygon-shaped cavity design with bi-axially textured YBa$_{2}$Cu$_{3}$O$_{7-x}$ superconducting tapes covering the entire inner wall.
Using a 12-sided polygon cavity, we obtain substantially improved quality factors of the TM$_{010}$ mode at 6.9\,GHz at 4\,K with respect to a copper cavity and observe no considerable degradation in the presence of magnetic fields up to 8\,T.
This corresponds to the first demonstration of practical applications of superconducting radio frequency technology for axion and other research areas requiring low loss in a strong magnetic field.
We address the importance of the successful demonstration and discuss further improvements.
\end{abstract}
\maketitle

\begin{center}
\begin{figure*}[t]
    \centering
    \includegraphics[width=1\textwidth]{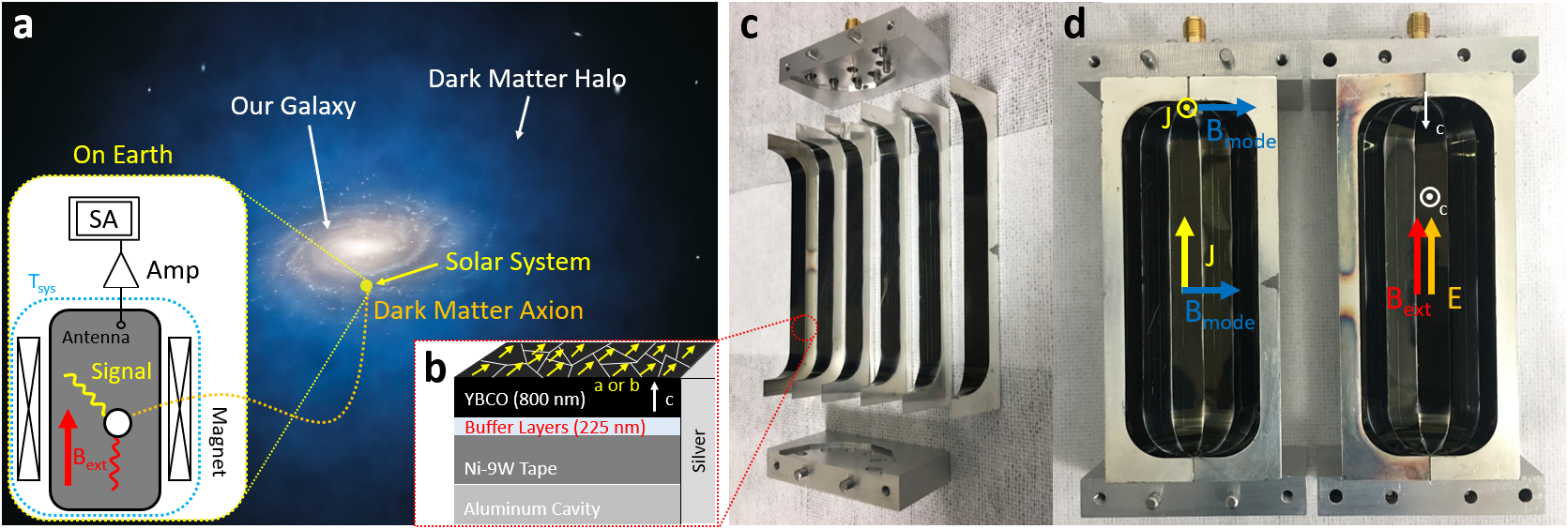}
    \caption{(a) A schematic of the axion haloscope illustrates the dark matter axion (dotted green line) in the dark matter halo (Credit: ESO/L. Calçada) converts to a GHz RF signal (curved yellow line) in the resonant cavity (gray box) via interaction with an external DC magnetic field (red arrow and curved red line, B$_{ext}$) from the magnet (white boxes with a cross). The cavity is installed in a cryogenic system (dotted blue box) with a system temperature (T$_{sys}$). The signal collected by antenna are detected by the spectrum analyzer (white box with "SA") after passing the amplifier (white triangle). Only a network analyzer is used in the Q factor measurement instead of the spectrum analyzer and the amplifier. (b) The schematic of the cross-section of a cavity piece shows the structure of the various layers (YBCO, Buffer Layers, Nickel-Tungsten) and the silver layer on the aluminum cavity. The c-axis (white arrow) of the YBCO crystal is parallel to the normal direction of the tape. In the biaxially textured YBCO film, the a or b axes (yellow arrows) of each single crystal YBCO grain are aligned to the tape direction. (c) Six aluminum cavity pieces; a YBCO tape is attached to each. (d) Twelve pieces (two cylinder halves) are assembled to make a whole cavity. The magnetic fields of a TM$_{010}$ mode (left, blue arrows) on the inner surface (black area) point in the vertical direction of the gap between the pieces. The surface current directions (left, yellow arrow and circle) are parallel to the tape direction. The c-axes of the YBCO grains (right, white arrow and circle) are vertical to the tape surface. At the middle of the cavity, the electric fields of the TM$_{010}$ mode (right, green arrow) which maximize the axion to photon conversion power are parallel to the external magnetic field (right, red arrow).}
    \label{fig:01}
\end{figure*}
\end{center}

\begin{figure}[t]
    \centering
    \includegraphics[width=0.45\textwidth]{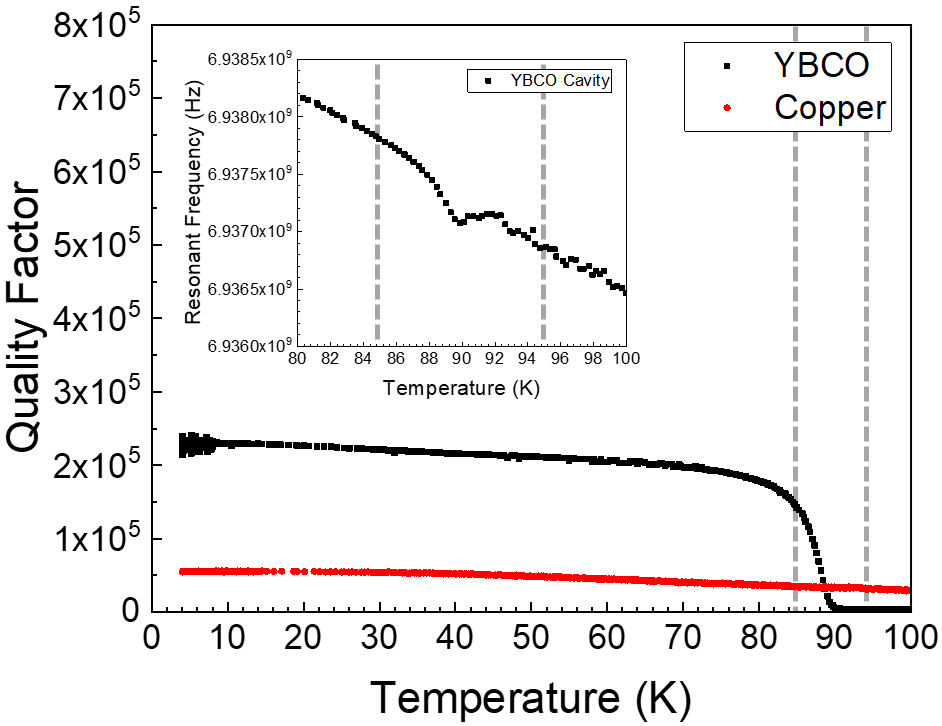}
    \caption{The measurement results of the 12-piece polygon cavities: The Q factor vs. temperature from 4.2\,K to 100\,K. The black dots are for the YBCO cavity and the red dots are for the copper cavity with the same polygon geometry. The inset plot is the resonant frequency vs. temperature from 80\,K to 100\,K. The phase transition from normal metal to superconductor starts near 90\,K, at which an anomalous frequency shift occurs. The vertical grey dashed lines show the temperatures 85\,K and 95\,K.}
    \label{fig:03}
\end{figure}

\begin{figure}[t]
    \centering
    \includegraphics[width=0.45\textwidth]{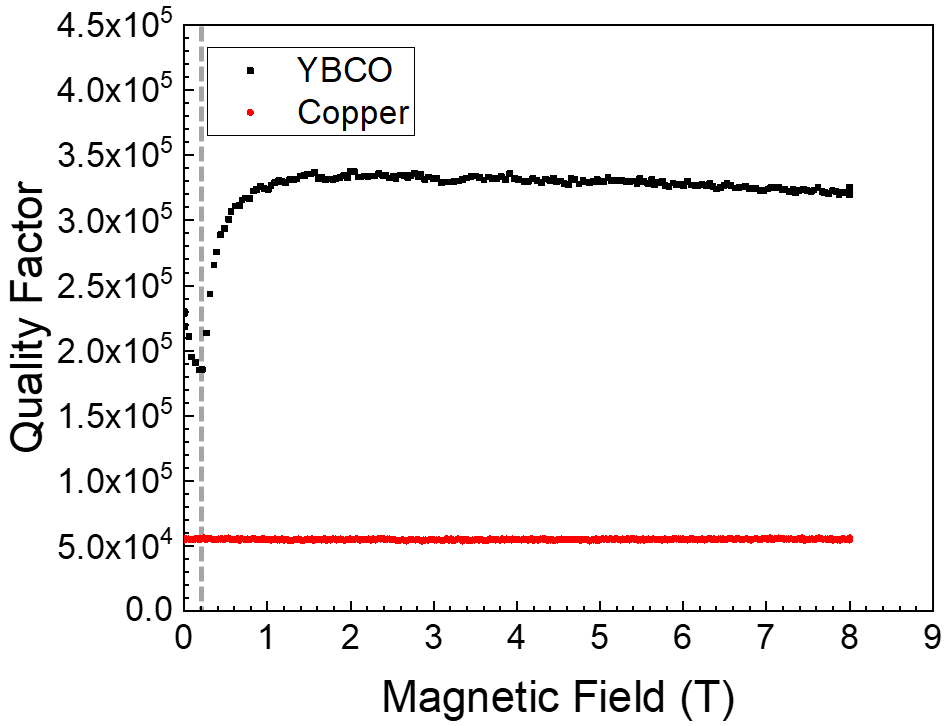}
    \caption{The measurement results of the 12-piece polygon cavities: The Q factor vs. external magnetic field from 0\,T to 8\,T. The vertical dashed line shows the magnetic field 0.23\,T at which the abrupt Q factor enhancement starts. The maximum Q factor is around 335,000.}
    \label{fig:04}
\end{figure}

\begin{figure}[t]
    \centering
    \includegraphics[width=0.45\textwidth]{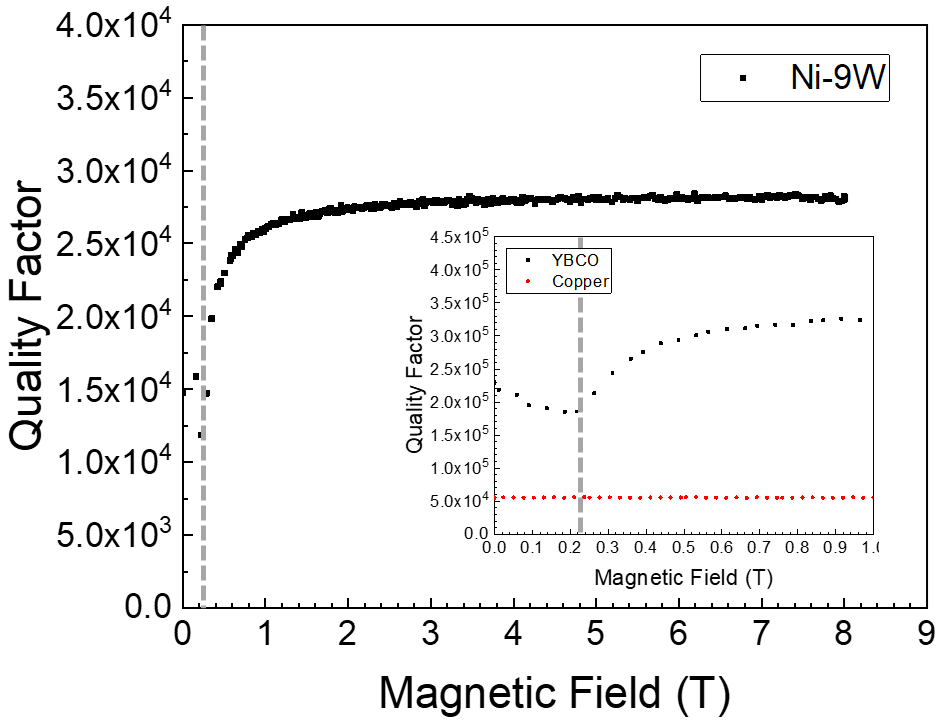}
    \caption{The measurement related to the nickel-tungsten alloy: The Q factor vs. external magnetic field for the 12-piece copper polygon cavity with one piece of Ni-9W (from 0 T to 8 T at 4 K). The Q factor behavior is the same as in the YBCO cavity. There is a abrupt change of the Q factor at 0.23 T.}
    \label{fig:05}
\end{figure}

\begin{figure}[t]
    \centering
    \includegraphics[width=0.45\textwidth]{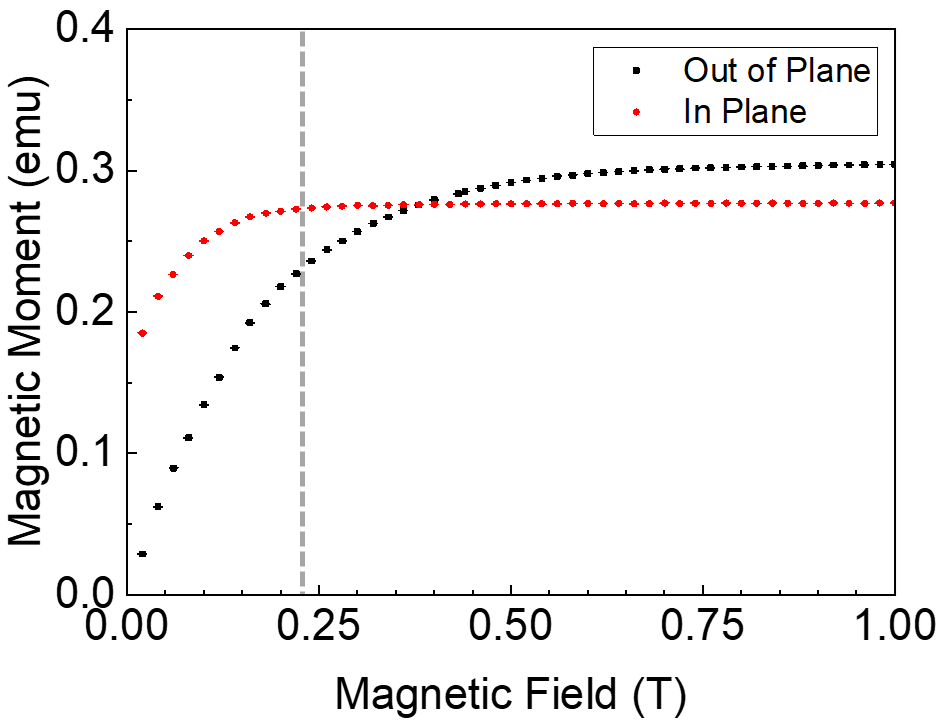}
    \caption{The measurement related to the nickel-tungsten alloy: The magnetization curve for the Ni-9W tape (4mm x 4mm). The magnetic saturation of the Ni-9W tape which is aligned in-plane direction of the DC magnetic field almost ends at 0.23 T at which the Q factor of YBCO cavity is abruptly changed. The vertical dashed line represents 0.23 T. For the case of out-of-plane, the magnetic saturation ends around 1 T at which the Q factor of the YBCO cavity is saturated.}
    \label{fig:06}
\end{figure}

The advancement of the superconducting radio-frequency (SRF) technology allows an RF resonator to obtain an extremely high quality (Q) factor and to be used in a broad scope of applications \cite{SCcavappli_Qubit_01, SCcavappli_Qubit_02, SCcavappli_Accel_01, SCcavappli_Accel_02}. However, the presence of an external magnetic field could limit scientific productivity in many areas where a strong external magnetic field is absolutely necessary. The examples include the beam screen design at the future circular collider (FCC) \cite{SCcavappli_Accel_03, SCcavappli_Accel_04} and high Q-factor cavities in axion dark matter research. In particular, the axion dark matter detection scheme proposed by P. Sikivie \cite{AxSearch_Sikivie01, AxSearch_Sikivie02} utilizes a resonant cavity immersed in a strong magnetic field ($>$\,8\,T) \cite{AxSearch_SCcavity_01, AxSearch_CavityExp_01, AxSearch_CavityExp_02, AxSearch_CavityExp_03, Hong14}, by which the axions are converted into microwave photons (Fig.\,\ref{fig:01}\,(a)).
Obtaining high Q factor in a strong magnetic field will profoundly impact the way axion dark matter experiments are performed.
It will substantially increase the axion to photon conversion power \cite{AxSearch_ScanRate} with expected quality factors of about 10$^{6}$ (Eq.\,\ref{eq:ax2ph}) \cite{AxSearch_AxionQfactor} and will permit investigations of the detailed axion signal structure in the frequency domain. Furthermore, achieving  a quality factor of more than $10^{8}$ can open a new window for ultra-narrow axion linewidth research \cite{AxSearch_NonVirialized}.\\
\indent Recently, the high frequency response of biaxially textured, second generation (2G) Rare-earth Barium Copper Oxide (REBa$_{2}$Cu$_{3}$O$_{7-x}$, REBCO) coated conductors (CCs) (Fig.\,\ref{fig:01}\,(b)) were studied in high magnetic fields up to 16\,T for the beam screen of the future circular hadron-hadron collider (FCC-hh) \cite{REBCO_Propert_01}.
For an ideal cylindrical cavity with a CC surface, the Q factor can be expected at least 10$^6$.
Considering the typical experimental conditions in dark matter axion search, the Q factor can be expected to be larger, because the vortex pinning becomes stronger at lower temperature (100\,mK) and in a magnetic field parallel to a REBCO film \cite{YBCO_Propert_Vortex_Golosovsky_01, YBCO_Propert_Vortex_Golosovsky_02, YBCO_Propert_Vortex_vertical}.
A high depinning frequency ($>$\,10\,GHz) is also reasonable for a dark matter axion search which target frequency ranges up to 100\,GHz.\\
\indent Fabricating a 3-D resonant cavity structure with 2G REBCO film poses large technical challenges because of its biaxial texture. Directly forming a biaxially textured REBCO film on the deeply concaved inner surface of the cavities is prohibitively difficult because of the limitations in making the well textured buffer layers and substrate \cite{YBCO_Fabrication_01, YBCO_Fabrication_02, YBCO_Fabrication_03, YBCO_Fabrication_04}. 
A possible solution to this problem is to implement a three-dimensional (3-D) surface with two dimensional (2-D) planar objects. We took advantage of high-grade, commercially available YBa$_{2}$Cu$_{3}$O$_{7-x}$ (YBCO) tapes from American Superconductor (AMSC), whose fabrication process, structure, and properties are well-known \cite{YBCO_Tape_01}.
The substrate and buffer layers of the tape were designed to act as template layers to provide a biaxial texture to the YBCO film. The film architecture of the tape consists of several parts. On the biaxially textured 9 percent nickel-tungsten (Ni-9W) alloy, the 800 nm thick YBCO was deposited on top of the buffer layers which consisted of Y$_2$O$_3$, YSZ, and CeO$_2$, and were each 75\,nm thick (Fig.\,\ref{fig:01}\,(b)).\\ 
\indent To fabricate a 3-D superconducting cavity utilizing YBCO tapes, we devised a novel scheme employing a 12-piece polygon cavity to which biaxially textured tap- 
\onecolumngrid
\twocolumngrid
\noindent es are attached. Each tape was prepared and attached securely to the inner surface of a cavity piece with minimum bending to prevent cracks (Fig.\,\ref{fig:01}\,(c)).
An arc radius of 10\,mm was applied between the top/bottom and the sidewall surfaces to avoid excess bending stress on the tapes \cite{YBCO_Tape_Bending_01}.
A joint mechanism for the twelve separated cavity pieces was designed for accurate alignment of the YBCO tapes upon assembly.
The fabrication and alignment error was within 20\,$\mu$m.\\
\indent In the axion haloscope, the fundamental TM modes which maximize an axion to photon conversion power are most commonly used. The axion to photon conversion power ($P_{a \gamma \rightarrow \gamma}$) is given by \cite{AxSearch_ScanRate},
\begin{equation}
\label{eq:ax2ph}
P_{a \gamma \rightarrow \gamma} = g^2_{a \gamma \gamma} {\rho_a \over m^2_a} B^2_{ext} V \omega_0 C {Q_c Q_a \over Q_c + Q_a}.
\end{equation}
The conversion power is related to the coupling constant ($g_{a \gamma \gamma}$), the density of dark matter axions ($\rho_a$), the rest mass of axions ($m_a$), the Q factor of the axion signal ($Q_a$\,$\sim$\,10$^6$), the volume average of a square of external magnetic field ($B^2_{ext}$), and the properties of a cavity which are the volume ($V$), the resonant frequency ($\omega_0$), the Q factor ($Q_c$), and the form factor ($C$). The form factor of a cavity for the given electric and electric displacement field of a resonant mode ($\mathbf{E}$,$\mathbf{D}$) (Fig.\,\ref{fig:01}\,(d)) and the vacuum magnetic susceptibility ($\mu_0$) is,
\begin{equation}
\label{eq:formfac}
C = {{\mu_0 \left( \int \mathbf{E} \cdot \mathbf{B_{ext}} dV \right)^2} \over {B^2_{ext} V \int \mathbf{D} \cdot \mathbf{E} dV}}.
\end{equation}
The COMSOL simulation \cite{Simulation_COMSOL} shows that the form factor of the prototype cavity is maximized by the TM$_{010}$ mode ($C$\,$\sim$\,0.67), because the electric fields are parallel to the external magnetic field along the vertical axis of the cavity. The external magnetic field distribution is given by the field map data of an 8 T cryogen-free NbTi superconductor solenoid \cite{Equipment_Magnet}. Also, in TM$_{010}$ mode, the primary surface loss originates from the side walls, where the surface resistance is minimized due to anisotropy of YBCO. At the sidewalls, the c-axis of YBCO film is perpendicular to the external magnetic field (Fig.\,\ref{fig:01}\,(d)). \\
\indent The vertical cuts in the cylindrical cavity do not cause any significant degradation of the Q-factor due to electromagnetic (EM) wave propagation through the gaps between the pieces.
In the simulation with perfect conducting surfaces, the cavity Q factor is maintained at 10$^8$ even with large (100\,$\mu$m) gaps between the pieces and tilted angle of gaps within the fabrication error ($<4\times10^{-4}$\,rad). The RF design cause no problem for high Q factor cavity up to 10$^8$.\\
\indent Once the YBCO tape was completely attached to the inner surface of each polygon piece, we removed the silver protective layers to expose the bare YBCO surface using a mixture of hydrogen peroxide and 60\,$\%$ ammonium water with 1:1 volume ratio. The cut edges of the YBCO tapes exposed on the side were coated by sputtering silver to reduce the RF loss due to the side surface of the layers under the YBCO film (Fig.\,\ref{fig:01}\,(b)). The technique used in this work was optimized for the TM modes of a cylindrical cavity but could be applied to any resonators, to minimize surface losses and resolve contact problems.\\
\indent The assembled cavity was installed in a cryogen-free dilution refrigerator BF-LD400 \cite{Equipment_BF}, equipped with an 8 T cryogen-free NbTi superconductor solenoid, and brought to a low temperature of around 4 K (Fig.\,\ref{fig:01}\,(a)). The Q-factor and resonant frequency were measured using a network analyzer through a transmission signal between a pair of RF antennae, which are weakly coupled to the cavity. The coupling strengths of the antennae were monitored throughout the experiment and accounted for in obtaining the unloaded quality factor. Measuring the Q factor (TM$_{010}$ mode) of the polygon cavity with the twelve YBCO pieces by varying the temperature, we observed the superconducting phase transition at around 90 K which is in agreement with the critical temperature (T$_c$) of YBCO (Fig.\,\ref{fig:03}). The global increase in the resonant frequency was due to thermal shrinkage of the aluminum cavity, but an anomalous frequency shift was also observed near the critical temperature. The decrease in the frequency shift at T$_c$ can be attributed to the divergence of the penetration depth of the YBCO surface \cite{YBCO_Propert_Vortex_Golosovsky_01}. The maximum Q factor at 4.2\,K was about 220,000. The Q-factor for the polygon cavity made of pure (oxygen-free high thermal conductivity, OFHC) copper with the same geometry was measured to be 55,500. Varying the applied DC magnetic field from 0\,T to 8\,T, at the initial ramping up of the magnet, the Q-factor of the cavity dropped rapidly to 180,000 until the magnetic field reached 0.23 T and then rose up to the maximum value of 335,000, which is about 6 times higher than that of a copper cavity, at around 1.5\,T for the TM$_{010}$ mode. From the measurement, we observed that the Q-factor of the resonant cavity's TM$_{010}$ mode did not decrease significantly (changing only a few percent) up to 8\,T (Fig.\,\ref{fig:04}).\\
\indent Investigating the abrupt behavior of the measured Q-factor near 0.23 T, the same Q-factor measurement was repeated using a 12-piece Cu cavity with Ni-9W tape attached to only one piece. The comparison between Fig.\,\ref{fig:04} and Fig.\,\ref{fig:05} clearly shows that the unexpected change in Q-factor near 0.23\,T was caused by the Ni-9W layer behind the YBCO layer. Furthermore, we also measured the magnetization of a small Ni-9W piece (4\,mm\,$\times$\,4\,mm) in a magnetic property measurement system Quantum Design MPMS3-Evercool \cite{Equipment_MPMS} with in-plane and out-plane alignments (parallel and perpendicular to the applied magnetic field, respectively) to investigate where the saturation occurs. We used nitric acid to etch the YBCO film on the 4\,mm\,$\times$\,4\,mm tape. The sample was installed in the equipment with a straw so that the rectangular sample could be aligned in any direction by hand. The measurements showed that the magnetic saturation of the in-plane (out-of-plane) Ni-9W ends near 0.23\,(1.0)\,T (Fig.\,\ref{fig:06}). The magnetic saturation lowers the surface resistance, because the atomic spins become more rigid due to the reduction of the magnetic domain walls.
In other words, the Q factor is suddenly increased at 0.23\,T because the main surface loss originates from the side wall, which is aligned in the in-plane direction with the external field. After that, the other surfaces are saturated until 1.0\,T, where the Q factor of the YBCO cavity is maximum.\\
\indent The maximum Q-factor achievable with a REBCO cavity is currently unknown but the comparison between the surface resistance of copper ($\sim$\,5\,m$\Omega$, 7\,GHz) and YBCO at 4\,K ($<$\,0.2\,m$\Omega$, 7\,GHz with an 8\,T parallel to the c-axis) \cite{REBCO_Propert_01, YBCO_Propert_Vortex_Golosovsky_01, YBCO_Propert_Vortex_Golosovsky_02, YBCO_Propert_Vortex_vertical} suggests that the Q-factor could be as much as 25 times higher than a copper cavity even with a strong magnetic field present. In the near future, improvements are expected using techniques to expose the bare YBCO surface from the tape, eventually reducing the area where surface losses occurs inside the cavity. Moreover, if the layer which creates the large energy loss, such as Ni-9W, can be eliminated or covered completely, we can expect a much higher Q factor. Our design of the vertically split, polygon cavity to implement the biaxially textured YBCO to on the inner surface allows us to test the potential of superconducting resonant cavities which could be used in a strong magnetic field. We have demonstrated that it is possible to fabricate a cavity with a YBCO inner surface to maintain a high Q-factor up to 8\,T. This result could not only eliminate a significant limitation of SRF applications with a magnetic fields in many areas, but also provide us with a tool to enhanced dark matter axion search.\\

\indent The authors are grateful for the technical advice of Sergey Uchaikin (Magnetic property of YBCO), Junu Jeong (Data Aquisition), Dongok Kim (General Discussion), Younggeun Kim (General Discussion) at the Center for Axion and Precision Physics Research in the Institute for Basic Science, and Byoungkook Kim at the KAIST Analysis Center for Research Advancement (Magnetic Property Measurement System). This work was supported by IBS-R017-D1-2021-a00 and IBS-R017-Y1-2021-a00 of the Republic of Korea.

\end{document}